\documentclass[
bibnotes,amsmath,amssymb,aps,
pra
]{revtex4-2}

\usepackage{graphicx}
\usepackage{dcolumn}
\usepackage{bm}
\usepackage{multirow}
\usepackage{amsmath}
\usepackage{color,soul}
\usepackage{hyperref}
\usepackage[dvipsnames,svgnames,table]{xcolor}
\newcommand{\ci}{\mathrm{i}}

\newcommand{\bea}{\begin{eqnarray}}
\newcommand{\eea}{\end{eqnarray}}
\newcommand{\bei}{\begin{itemize}}
\newcommand{\eei}{\end{itemize}}
\newcommand{\be}{\begin{equation}}
\newcommand{\ee}{\end{equation}}
\newcommand{\bse}{\begin{subequations}}
\newcommand{\ese}{\end{subequations}}
\newcommand{\bfg}{\begin{figure}}
\newcommand{\efg}{\end{figure}}

\newcommand{\uvec}[1]{\boldsymbol{\hat{\textbf{#1}}}}

\begin{document}

\preprint{APS/123-QED}

\title{Effect of the excitation setup in the improved enhancement factor of covered-gold-nanorod-dimer antennas}

\author{Iván A. Ramos\textsuperscript{1,2}, L. M. León Hilario\textsuperscript{1}, María L. Pedano\textsuperscript{2}}
\author{A. A. Reynoso\textsuperscript{2,3}}
\altaffiliation[Email: ]{reynoso@cab.cnea.gov.ar}
\affiliation{\textsuperscript{1}Facultad de Ciencias, Universidad Nacional de Ingenieria, Apartado 31-139, Av. Túpac Amaru 210, Lima, Perú\\
\textsuperscript{2}Centro At\'omico Bariloche and Instituto Balseiro, CNEA, CONICET, 8400 San Carlos de Bariloche, R\'io Negro, Argentina\\
\textsuperscript{3}Departamento de F\'isica Aplicada II, Universidad de Sevilla, E-41012 Sevilla, Spain}

\date{9.~July 2022}

\begin{abstract}
Devices possessing the ability to sense both electrically and optically molecular targets are of fundamental and technological interest. Towards this end, it has been shown that covering the ends of gapped gold-nanorod-dimer nanoantennas can improve the enhancement factor (EF) that quantifies the nanoantenna efficiency for surface-enhancement Raman spectroscopy (SERS) for an incident wave coming from the top of the sample. Here, as the covering breaks the top-bottom symmetry, we investigate the behavior of the EF for excitation coming from the bottom of the sample. This is relevant in presence of a reflecting substrate or due to the placement of the device in a cavity field. We also study the case of a superposition of waves coming from both directions in the limit cases in which a node or an antinode of the total incident field lies at the center of the gold nanorods. In all these situations we find that the EF of the covered device can continue to be higher than for the uncovered case when the geometrical parameters are tuned to the peak values of the calculated enhancement factor. 
\end{abstract}
\maketitle

\section{Introduction}

In the last decades, a synergy between theory and experimental advancements in the field of nanoplasmonics, \cite{StockmanREVIEW}   has led to a great deal of research leading to technological applications in energy, sensing and other fields \cite{Kurt2021,ChenRSCadv2022,ZhaoRSCadv2022}. Among the different platforms, metallic nanoantennas forming dimers are designed to present a hot-spot of the local electric field at the region between its two  constituents, i.e., at the nanogap. This effect is seen in several different geometries as for example bowtie type or rectangular nanoantennas synthesized by electron beam microscopy \cite{novo,ML5-BraunNanophotonics2018,ML7-PucciPSS2010}. The effect involves collective charge oscillations known as surface plasmons \cite{ML1-JainJPC2008,ML2-FunstonNanoLett2009} being constrained by the geometry of the sample for the particular incoming excitation. The hot-spots are associated to localized surface plasmon resonances (LSPRs)  \cite{ML3-GianniniChemicaReview2011} that allow for efficient application of surface enhanced Raman spectroscopy (SERS) \cite{zhen,ML18-Osberg2012,AlexanderNanoLet2010,slau2010}. A wide range of technological applications become available by identifying in the SERS signal characteristic vibrations of target molecules that lie inside the nanogap \cite{zhen,cao,hutt}.  

\begin{figure}
\includegraphics[width=0.5\columnwidth]{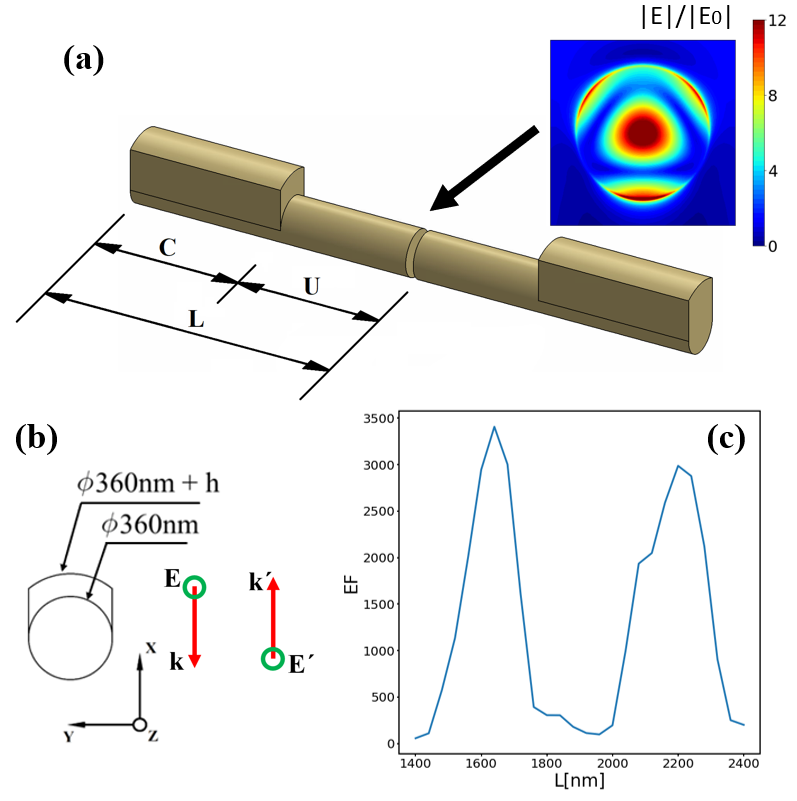}
\centering
\caption{\label{fig:1} (a) Scheme of a covered gold-nanorod-dimer antenna and definition of its length-related geometrical parameters. (b) Covered part's profile indicating the nanorod diameter and the thickness of the cover layer, $h$. The antenna is mono-chromatically illuminated by a plane wave of wavevector $\mathbf{k}$, $\lambda=633$nm, and having electric field polarization along the nanorod axis $z$. (c) Enhancement factor as function of $L$ is shown for the case of $\uvec{k}=-\uvec{x}$, $C=1200$nm, and $h=200$nm ---the calculated nanoantenna EF arises from the amplitude of the electric field at the gap (see inset of  panel (a)). In this work we also consider the case of $\uvec{k}=+\uvec{x}$.}
\end{figure}

A prominent research direction in this field involves electrically contacting both parts of the nanoantenna dimer in order to enable electrical transport measurements through the molecules at the gap. A successful design of the contacts must not degrade the nanonantenna enhancement factor (EF) \cite{pablitoE4,librochapter5}, namely, a figure of merit for SERS spectroscopy that averages the fourth power of the local electric field amplitude over a surface inside the nanogap. In addition, the ability of the original dimer to contain a large number of molecules improves the electrical signal. Recently, in Ref.\cite{IvanRSC2021}, a promising result in this direction have been reported for gold-nanorod-dimer nanoantennas as they theoretically reach larger values of the EF when a section of the edges of the dimer are covered by a gold layer. Importantly, as the diameter of the opposing nanorods can be hundreds of nanometers the gap is able to host a large number of molecules that would lead to a robust electrical signal. 

The uncovered version of gold-nanorod-dimer antennas was fabricated and measured in Ref.\cite{pedano}. The structure of the local electric field arises due to interaction between the incoming light and surface plasmons generating surface plasmon polaritons (SPPs) constricted to the particular geometry. In these antennas the local field at the nanogap is found to be maximized for incoming-wave-electric field polarized parallel to the nanorod axis and, therefore, the incident wavevector is normal to the nanorod axis. Fixing the nanorods' diameter and gap distance the SERS intensity was found to develop a periodic dependence on the length of each gold nanorod. Peak situations are reached at particular geometrical parameters maximizing the nanogap local field, precisely at the maxima of the calculated EF \cite{shuzho} signaling the presence of hot-spots. Importantly, as the obtained EF peak values remain high for several micrometers-long it becomes technologically feasible to add extended covered regions far from the gap. These covered regions can be addressable electrodes \cite{ML16-Chen2008,ML17-Lim2016}, and electric transport measurements can be added to the SERS spectroscopy \cite{chen,ML19-Chen2009,ML6-FlorianNanophotonics2020}. This type of cover designs have been theoretically investigated in Ref.\cite{IvanRSC2021} where it was shown that, by controlling the covered dimmer's geometrical parameters, the EF can reach larger values than for the uncovered dimer antenna. This improvement, which arises as a result of a geometrical-induced rearrangement of the surface plasmon polaritons, was shown to be robust to the presence of dimer asymmetries and vacancies at the interfaces between the nanorods and the covering layers.

In the Ref.\cite{IvanRSC2021} the incident wave comes from the top of the sample. It must be noted, however, that the cover layers break the top-bottom symmetry of the sample. This motivates the study of the enhancement factor behaviour due to incident waves coming from the bottom of the sample; in practice this wave may be originated from the reflection at the substrate. In this work, we consider the case in which the incident wave comes from the bottom of the sample in order to understand the greatest effect that the top-bottom-symmetry breaking can produce on the rearrangement of the SPPs and thus on the conditions to generate hot-spots. In both instances, the polarization of the incoming wave is kept along the nanorod axis as this maximizes the local field at the nanogap. We identify that the optimal length of the dimer changes differently in the two situations. Finally, we consider the superposition of both excitation directions in two limiting cases, namely, the even and odd combination with respect to the top-bottom coordinate. The analysis allows us to conclude that when the presence of the reflected wave is important---either due to a highly reflecting substrate or if the device were placed inside a cavity field---the preferred design choice corresponds to the parameters that optimize the excitation coming from the bottom direction. This result has practical implications for the design of the antennas and it further demonstrates that the geometry should be optimized to the particular substrate conditions, or cavity arrangement, in which it will be operated.   

The paper is organized as follows: In Sec.\ref{SC:Model} we introduce the sample design, describe the numerical modelling and provide the definitions of the enhancement factor and the location-resolved average charge densities. In Sec.\ref{SC:Results} we present and compare the results of enhancement factor for all the excitation configurations mentioned above. In Sec.\ref{SC:Conclusions} we conclude and discuss the main findings. 

\section{Definition and modeling of covered gold-nanorod-dimer nanoantennas}
\label{SC:Model}
\subsection{Covered dimer geometry}
Cylindrical gapped gold nanoantennas can be fabricated by on-wire lithography (OWL) \cite{quin,ML4-Mangelson2013,ML14-Hurst2006},  this technique achieves sub-5nm fabrication resolution \cite{ML15-Braunschweig2010} involving the electrochemical deposition of metals using the pores of anodic aluminum oxide (AAO) templates of fixed diameter. In Refs.\cite{pedano,shuzho} the choice of 360nm nanorod diameter was found to produce gold nanorods with excellent quality reaching length of several micrometers. Nanorods with smaller diameter values had a tendency to break disallowing the possibility to add wide contacted regions. With that in mind, and targeting the detection of molecules such as DNA, we consider a gap distance of $25$nm, fix the nanorod diameter in 360nm, and adopt the design of covered nanoantenas presented in Ref.\cite{IvanRSC2021} where alternative synthesis and covering procedures were discussed \cite{EduardoX1-3-Jiu2014,EduardoX2-Martinez_2016,EduardoX3-1-Si2013,EduardoX3-2-Jiang2015}.

In Fig.\ref{fig:1}(a) we present the sketch of the design: a cover layer of height $h$ has been added far from the gap to both nanorods (aligned to the $z$-axis) composing the dimer. Each nanorod has length $L$ and a covered region of length $C$, thus the uncovered region has length $U=L-C$. By studying the EF dependence on $L$, $U$, $C$ and $h$ we can readily compare the performance of the antenna with respect to the experimentally validated and feasible setup of the uncovered dimer of Refs.\cite{pedano,shuzho}. This comparison is direct because we use the diameter, wavelength (see below) and gap distance given in the mentioned references and the figure of merit of the antenna, EF, involves the very same computation that the uncovered case. However, the main conclusions are not limited to this case and can be readily extended to other nearby excitation wavelengths, gap distances and nanorod diameters. 

\subsection{Optical Excitation Setup}
As helium–neon continuous-wave lasers are widely available we fix the excitation wavelength at $633$nm. This choice also allows us to compare the EF here obtained with the ones of previous works \cite{shuzho,IvanRSC2021}. In Fig.\ref{fig:1}(b) we show the wavevector, $\mathbf{k}$, and the polarization of the excitation. In order to maximize the electric field at the nanogap, 
we assume two different wavevector direction, $\uvec{k}=\pm \uvec{x}$, and always perpendicular to the $z$-axis and the electric field is polarized along $z$-axis \cite{pedano,shuzho}.

The geometry of the covering, being even along the $y$-axis, generates a sample which is no longer symmetrical with respect to the $x=0$ plane thus breaking the top-bottom symmetry lying along the $x$-direction. As mentioned in the introduction, in this work we study the case in which the incident waves comes from the bottom of the sample, (i.e., $\uvec{k}=+ \uvec{x}$). We compare the latter situation with extended results of the case of incident wave coming from the top of the sample (i.e., $\uvec{k}=- \uvec{x}$) that, in Ref.\cite{IvanRSC2021}, was shown to develop improved EF than the uncovered case. Finally, as is detailed in Sec.\ref{SC:Results}, we study the EF of the covered design when the system is subject to top-bottom even and odd combinations, namely, an applied oscillating electric field having either a maximum (antinode) or a node at $x=0$.   

\subsection{Enhancement factor and surface charge densities}
In its simplest description, the efficiency of SERS is locally characterized by the enhancement factor \cite{pablitoE4,librochapter5}, $|\mathbf{E}|^4/E_0^4$, where $\mathbf{E}$ and $E_{0}$ are the local electric field and the amplitude of the incident electric field, respectively. The EF for the full nanogap arises by averaging the local factor over a surface at that region:
\begin{equation}
EF=\frac{\int|\mathbf{E}|^4/E_0^4dS}{\int dS},
\end{equation}
with $dS$ the surface area differential over a plane parallel to one of the rod faces locate $2$nm inside the gap. 

Following the method of Ref.\cite{IvanRSC2021} we obtain the surface charge density $\sigma(\mathbf{r}_s)$ by applying at each surface position $\mathbf{r}_s$ the Gauss law. Then, for the characterization of the plasmonic $z$-dependence along each part of the dimer, we average, at each $z$, the surface charge density along the antenna section. We split this average surface charge density into the top and bottom part as given by     
\be
\tilde{\sigma}_{\pm}(z)=\frac{\int_{C_{z,\pm}}\sigma~\! d\ell }{\int_{\substack{C_{z,\pm}}} d\ell},
\ee
where $d\ell$ is the infinitesimal arc length and $C_{z,+}$ ($C_{z,-}$) is the curve along the perimeter of the nanoantenna at a given $z$ for the top (bottom) part, namely, for $x>0$ ($x<0$). This top and bottom splitting of the analysis is motivated by the fact that the diameter is not negligible with respect to the surface plasmon wavelength. In particular, $\tilde{n}_{{}_L \pm}$ ($\tilde{n}_{{}_U \pm}$), the number of nodes presented by $\tilde{\sigma}_{\pm}(z)$ along the full length (covered length) of a nanorod, has been shown to be useful to identify peak and valley differences on the arrangement of the SPPs with and without the presence of the cover layer \cite{IvanRSC2021}. In Sec.\ref{SC:Results} below, these numbers, specially $\tilde{n}_{{}_L -}$, prove to be useful resources in the interpretation of the results.   

\subsection{Numerical simulation details}
Several approaches to simulations in this area exist. Examples are: Fabry-Perot resonances in a single cylindrical wire including thin metal nanowires \cite{ML8-Gordon2009,ML9-Dorfmuller2009}, mode analysis optimizing complex plasmonic nanoantennas \cite{ML10-Dey2020}; integral equations formulation via the Method of Moments in array of nanorods \cite{ML11-Soliman2014}; finite-difference time-domain (FDTD) software (Lumerical) applied to the near-field of cascaded nanorod antennas \cite{ML12-Zhang2019}; a circuit equivalent of a
plasmonic nanoantenna \cite{ML13-Eggleston2015}, among many others. The discrete dipole approximation (DDA) method \cite{dra,drai,go} has been also extensively used and applied for this case in Ref.\cite{IvanRSC2021}. 

Here instead, the Maxwell equations are solved with the finite elements method. A 3D system is defined and a mesh is generated. This mesh, that includes both the sample and air, is carefully designed in order to properly describe the skin-depth effect within the gold and the plasmon profile immediately above the surfaces of the nano-antenna. For the largest simulated antennas the number of unknowns, i.e., the degree of freedom, reached $6\times 10^6$. In a typical up to day workstation solving the corresponding sparse matrix describing the system required up to $200$GB of memory, this was after imposing symmetry considerations that manage to reduce the computational cost of the calculation. The simulation times can be of up to several days for the sweeping of a single parameter. The enhancement factor follows from post-processing the local electric field within the gap obtained from full wave simulations having the sample subject to the desired excitation at a fixed wavelength, i.e., in the frequency domain. This was implemented using the software package COMSOL multiphysics which is extensively applied to several types of plasmonic devices \cite{PLASMON:comsol,COMSOL:nanomaterials2021}. We first verified good qualitative and quantitative agreement with simulations reported in previous works \cite{shuzho,IvanRSC2021}, that were obtained via the discrete dipole approximation method as implemented in the DDSCAT library \cite{dra}. 

\section{Results}
\label{SC:Results}
\begin{figure}
\includegraphics[width=0.9\columnwidth]{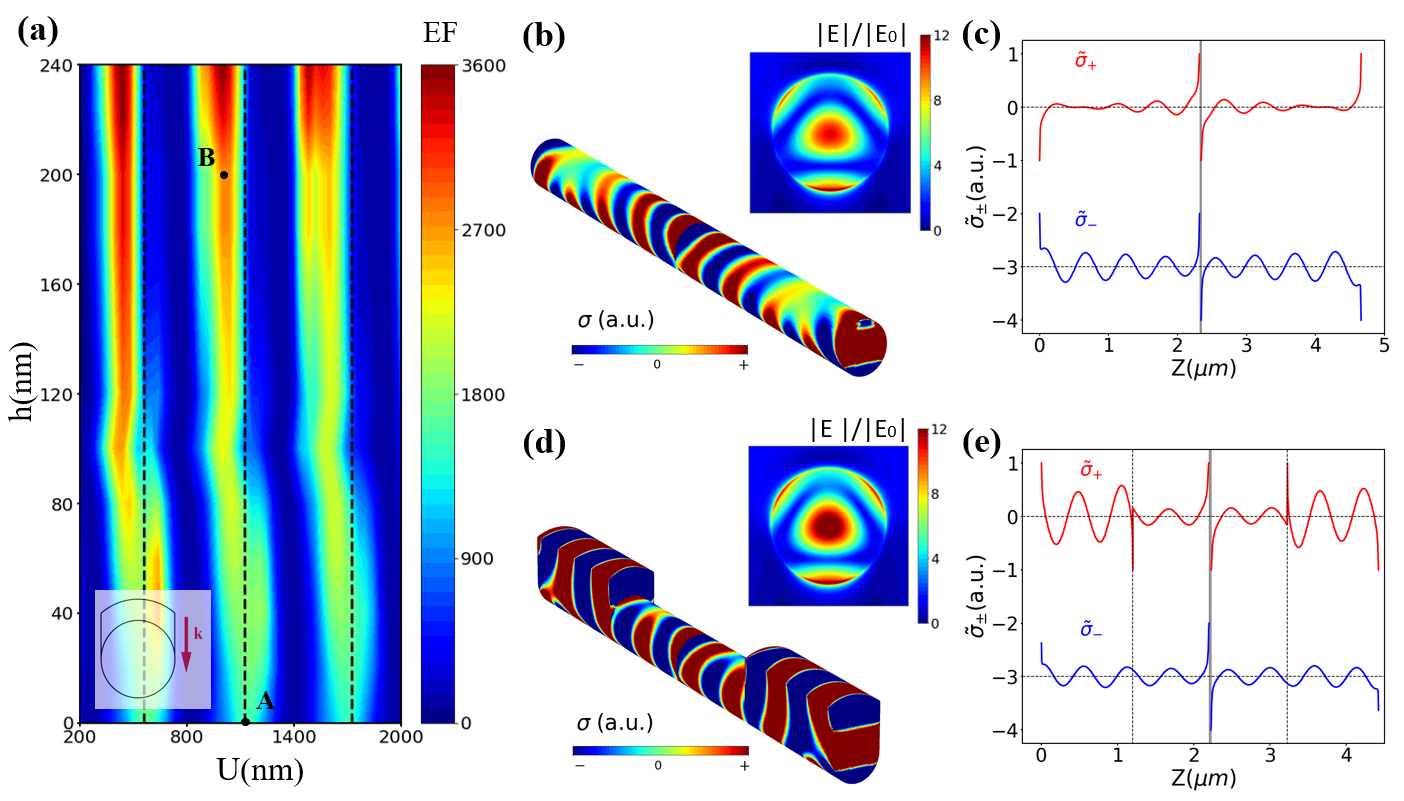}
\centering
\caption{\label{fig:2} Enhancement-factor colormaps as a function of $L$ and $h$ for $C=1200$nm and a plane wave driving with $\uvec{k}=- \uvec{x}$. Dotted lines indicate the $L= L_{0,i}$ conditions, i.e., the nanorod's lengths associated to peaks in EF for the uncovered dimer. The position of the peaks are shifted to shorter total length than the closest $L_{0,i}$ condition for the $\uvec{k}=- \uvec{x}$ incoming wave. (b) and (d) Surface charge density for peak $A$ and $B$ of uncovered and covered dimer, respectively. Colormaps of $|E|$ at the gap for peak $A$ and $B$ are included as insets. (c) and (e) represents the top, $\tilde{\sigma}_+(z)$, and bottom, $\tilde{\sigma}_-(z)$, average surface charge densities for peak $B$ and $A$ of the uncovered and covered dimer, respectively.
}
\end{figure}

We start with the incident wave coming from the top of the sample, this corresponds to taking $\uvec{k}=- \uvec{x}$ in Fig.\ref{fig:1}(b) and therefore it revisits the results reported in Ref.\cite{IvanRSC2021}. The typical enhancement factor as a function of the length $L=U+C$ is shown in Fig.\ref{fig:1}(c) for $h=200$nm and $C=1200$nm. In Fig.\ref{fig:2}(a), keeping $C$ fixed, we show the EF as a function of the total length $L$ and the covering thickness $h$. Vertical dashed lines show the values of the length, $L_{0,i}$, for which the EF has peaks in the original uncovered dimer, i.e., for $h=0$. Above a certain value of $h$ the EF has peaks at values of $L_i$ smaller than the closest $L_{0,i}$. For the two points marked in the colormap of Fig.\ref{fig:2}(a), the pattern of the SPPs can be visualized from the simulated surface charge density presented in Fig.\ref{fig:2}(b). 

Figure \ref{fig:2}(c) shows the associated averaged top-bottom resolved surface charge densities as a function of $z$. By analyzing $\tilde{\sigma}_{\pm}(z)$ at several EF peaks, for the covered design subject to the $\uvec{k}=- \uvec{x}$ incoming wave, one finds that the SPP patterns at EF peaks have $(\tilde{n}_{{}_U +},\tilde{n}_{{}_U -})=(2n,2n)$, i.e., both are even and $|\tilde{n}_{{}_U -} -\tilde{n}_{{}_U +}|=0$ \footnote{In the computation of $\tilde{n}_{{}_L +}$ we omit the nodes arising due to the sharp variation of $\tilde{\sigma}_{+}(z)$ found in the cover region near the position where the uncovered region begins. This is justified for an analysis that considers nodes if they can be associated, approximately, to the presence of an additional half wavelength of the SPP, a condition which is not met by the mentioned sharp feature in $\tilde{\sigma}_{+}(z)$.}. In contrast, one finds $(\tilde{n}_{{}_L +},\tilde{n}_{{}_L -})=(2n-1,2n)$ and $|\tilde{n}_{{}_L -} -\tilde{n}_{{}_L +}|=1$ for the associated (closest in length) EF peak of the uncovered dimer \footnote{For the uncovered design, because the top-bottom symmetry is not broken, incoming wave coming from the top or bottom of the sample leads to the same result: one only needs to interchange $\tilde{n}_{{}_L +}$ and $\tilde{n}_{{}_L -}$.}. This is consistent with finding these peaks of EF at different total length $L$ in the two cases.

\begin{figure}
\centering
\includegraphics[width=0.9\columnwidth]{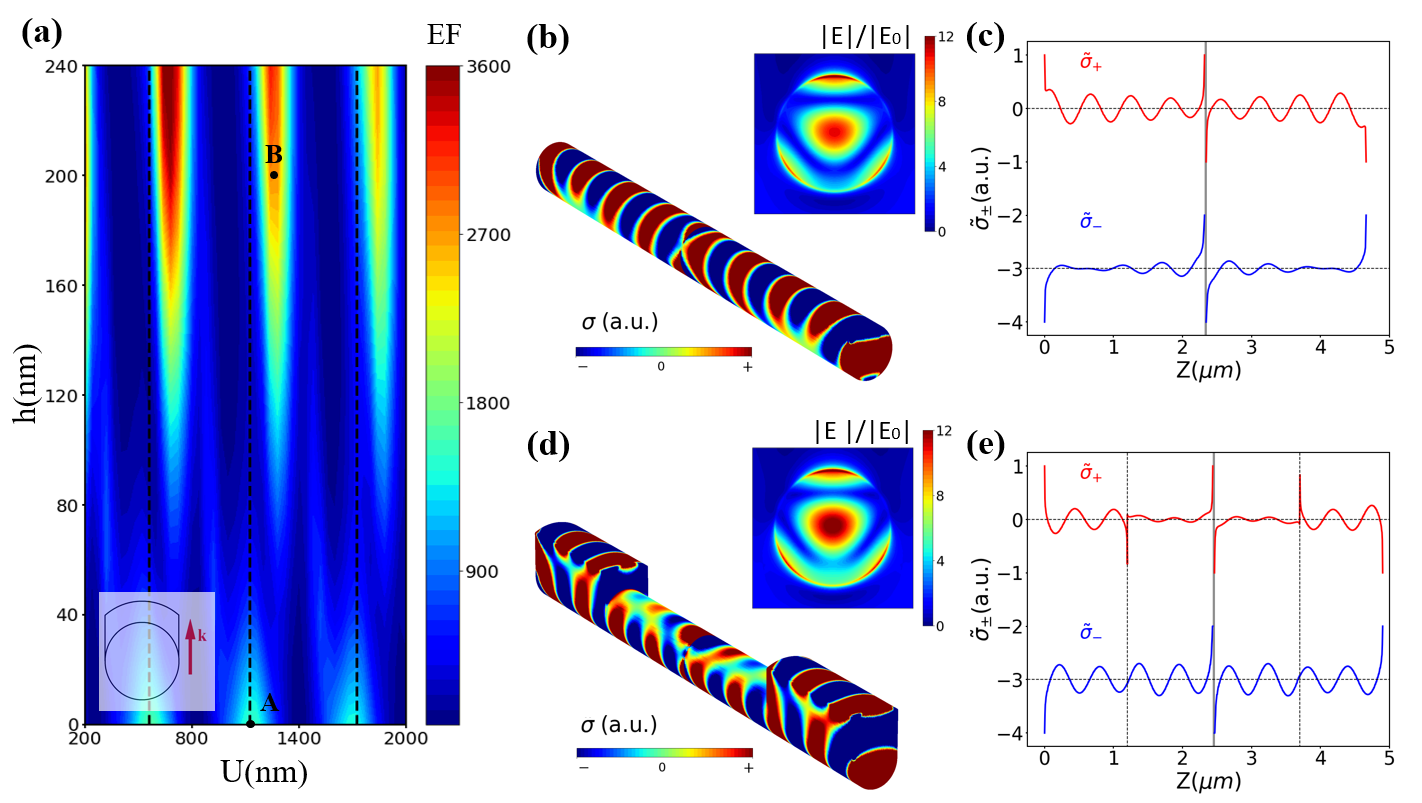}
 \caption{\label{fig:3} Enhancement-factor colormaps as a function of $L$ and $h$ for $C=1200$nm and a plane wave driving with $\uvec{k}=+ \uvec{x}$. As in the case of $\uvec{k}=- \uvec{x}$, here the dotted lines indicate the $L= L_{0,i}$ conditions. The position of the peaks are shifted to longer total length than the closest $L_{0,i}$ condition for the $\uvec{k}=+ \uvec{x}$ incoming wave. (b) and (d) Surface charge density for peak $A$ and $B$ of uncovered and covered dimer, respectively. Colormaps of $|E|$ at the gap for peak $A$ and $B$ are included as insets. (c) and (e) represents the top, $\tilde{\sigma}_+(z)$, and bottom, $\tilde{\sigma}_-(z)$, average surface charge densities for peak $A$ and $B$ of the uncovered and covered dimer, respectively. 
}
\end{figure}
\begin{table}
\caption{\label{tab:table1}Number of nodes along the uncovered part, $\tilde{n}_{{}_U \pm}$, for the uncovered and covered configuration (with fixed $h=200$nm and $C=1200$nm) considering either $\uvec{k}=-\uvec{x}$ or $\uvec{k}=+\uvec{x}$.}
\begin{center}
\begin{tabular}{cccc|ccc|cccc}
\multicolumn{4}{c|}{Covered, $\uvec{k}=-\uvec{x}$} & \multicolumn{3}{|c|}{Uncovered, $\uvec{k}=-\uvec{x}$ ($\uvec{k}=+\uvec{x}$)} &\multicolumn{4}{c}{Covered, ($\uvec{k}=+\uvec{x}$)} \\
\hline
  $L[nm]$ & $U[nm]$ & $\tilde{n}_{{}_U +}$ & $\tilde{n}_{{}_U -}$ & $L[nm]$ & $\tilde{n}_{{}_L +}$ & $\tilde{n}_{{}_L -}$ & $L[nm]$ & $U[nm]$ & $\tilde{n}_{{}_U +}$ & $\tilde{n}_{{}_U -}$ \\
  1640 & 440 & 2 & 2 & 640  & 1(2) & 2(1) & 1880 & 680 & (2) & (3)\\
  2200 & 1000& 4 & 4 & 1200 & 3(4) & 4(3) & 2440 & 1240& (4) & (5)\\
  2800 & 1600& 6 & 6 & 1760 & 5(6) & 6(5) & 3040 & 1840& (6) & (7)
\end{tabular}
\end{center}
\end{table}

\begin{table}
\caption{\label{tab:table2}Number of nodes along the full length, $\tilde{n}_{{}_L -}$, for the uncovered and the covered configuration (with fixed $h=200$nm and $C=1200$nm) considering either $\uvec{k}=-\uvec{x}$ or $\uvec{k}=+\uvec{x}$.}
\begin{center}
\begin{tabular}{cc|cc|cc}
\multicolumn{2}{c|}{Covered, $\uvec{k}=-\uvec{x}$} & \multicolumn{2}{|c|}{Uncovered, $\uvec{k}=-\uvec{x}$ ($\uvec{k}=+\uvec{x}$)} &\multicolumn{2}{c}{Covered, ($\uvec{k}=+\uvec{x}$)} \\
\hline
  $L[nm]$ & $\tilde{n}_{{}_L -}$ & $L[nm]$ & $\tilde{n}_{{}_L -}$ & $L[nm]$ & $\tilde{n}_{{}_L -}$\\
  1640 & 6 & 1760 & 6 (5)& 1880 & (7)\\
  2200 & 8 & 2320 & 8 (7)& 2440 & (9)\\
  2800 & 10& 2920 & 10(9)& 3040 & (11)
\end{tabular}
\end{center}
\end{table}
We proceed to investigate the case of $\uvec{k}=+ \uvec{x}$, i.e., with incident wave coming from the bottom of the sample. Figure \ref{fig:3} shows EF colormaps as a function of the total length $L$ and the covering thickness $h$. In this case above a certain value of $h$ the EF has peaks at values of $L_i$ larger than the closest $L_{0,i}$ position, marked with vertical dashed lines. Interestingly, this means that at fixed $C$ (and in general as we show below), the covered design develops its main EF-peaks at lengths that are larger or smaller than the uncovered case depending on the top/bottom orientation of the incoming wave. We perform a qualitative characterization of the SPP pattern for the $\uvec{k}=+ \uvec{x}$ case by analyzing $\tilde{\sigma}_{\pm}(z)$ at several EF peaks (as the one presented for $h=200$nm in Fig.\ref{fig:3}(c) which is derived by post-processing the $\sigma(\mathbf{r}_s)$ shown on Fig.\ref{fig:3}(b)) finding that the SPP patterns at these EF peaks have $(\tilde{n}_{{}_U +},\tilde{n}_{{}_U -})=(2n,2n+1)$, i.e. $|\tilde{n}_{{}_U -} -\tilde{n}_{{}_U +}|=1$ where the odd side corresponds to the $\tilde{n}_{{}_U -}$ values. In Table \ref{tab:table1} we present the $\tilde{n}_{{}_U \pm}$ values calculated considering $\uvec{k}=- \uvec{x}$ (left) and $\uvec{k}=+ \uvec{x}$ (right) for covered nano-antennas with fixed $h=200$nm and $C=1200$nm at several EF peaks. For the uncovered case, the table presents $\tilde{n}_{{}_L \pm}$ for EF peaks taken at the values $L_{0,i}$ that are closer to the $U$ values of the covered cases being compared at each row. Note that for $\uvec{k}=- \uvec{x}$ the $\tilde{n}_{{}_{U/L} -}$ are even whereas for $\uvec{k}=+ \uvec{x}$ the $\tilde{n}_{{}_{U/L} +}$ are odd. In addition to this, when the number of nodes is even they are equal to $\{2,4,6\}$ for all cases: $\tilde{n}_{{}_{L}-}$ ($\tilde{n}_{{}_{L}+}$) and $\tilde{n}_{{}_{U}-}$ ($\tilde{n}_{{}_{U}+}$) for $\uvec{k}=-\uvec{x}$ (for $\uvec{k}=+ \uvec{x}$). This peculiarity appears on the nodes at the opposite side of the face where the incident electric field wave enters.

\begin{figure}
\centering
\includegraphics[width=0.9\columnwidth]{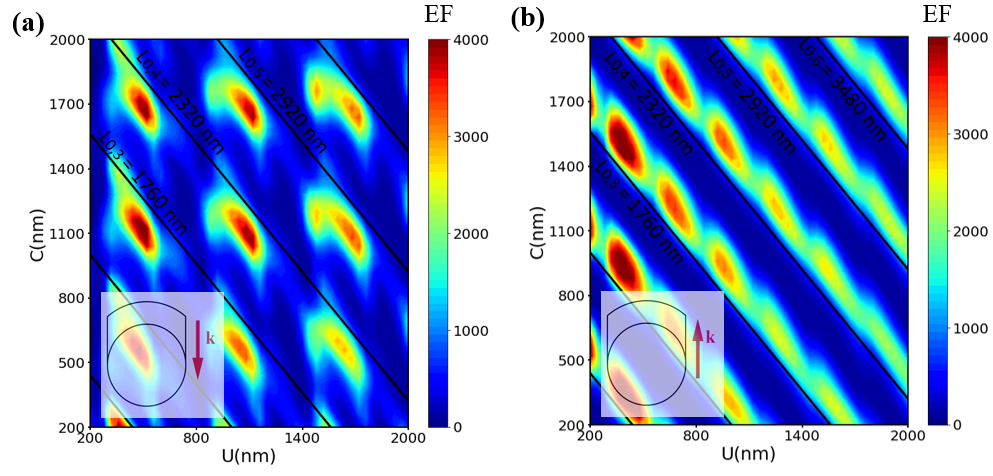}
\caption{\label{fig:4} Enhancement-factor colormaps as a function of $U$ and $C$ for $h=200$nm and a plane wave driving with (a) $\uvec{k}=-\uvec{x}$ ((b) $\uvec{k}=+\uvec{x}$). Solid lines indicate the $U+C= L_{0,i}$ conditions, i.e., the nanorod's lengths associated to peaks in EF for the uncovered dimer. Note that the maximum of EF at $P_{A}$ ($P_{B}$) is found shifted to shorter (longer) total length than the closest $L_{0,i}$ condition. }
\end{figure}

Additionally, Table \ref{tab:table2} includes the $\tilde{n}_{{}_L -}$ for the covered antennas at EF peaks. Similar to Table \ref{tab:table1}, the parity is even for $\uvec{k}=-\uvec{x}$ and odd for $\uvec{k}=+\uvec{x}$. The upper surface has not been taken into account due to the discontinuity in the covered-uncovered interface ($z=C$).  In both tables, a shift (in $L$) of the covered dimers peaks is observed with respect to the lengths producing EF peaks in the uncovered dimers. For $\uvec{k}=-\uvec{x}$ and $\uvec{k}=+\uvec{x}$ the peaks moved to lower and higher lengths, respectability. This displacement of the optimal lengths is also observed in Figure \ref{fig:4} for several values of $C$ in EF colormaps as a function of $U$ and $C$ for $\uvec{k}=-\uvec{x}$ in Fig.\ref{fig:4}(a) and for $\uvec{k}=+\uvec{x}$ in Fig.\ref{fig:4}(b). The peaks for the uncovered dimer are represented by solid lines ($U+C=L_{0,i}$) with slope of $-1$. For $\uvec{k}=-\uvec{x}$ we observe that there are some vertical bands, i.e., ranges in $U$, where no EF are found for any $C$ values. On the other hand, in the case of $\uvec{k}=+\uvec{x}$, one cannot easily identify ranges of $U$ having large degradation of the EF values. Importantly, for both directions, Fig.\ref{fig:4} the values of the EF peaks are larger than the maximum achieved EF in the uncovered device.

\begin{figure}
\centering 
\includegraphics[width=0.9\columnwidth]{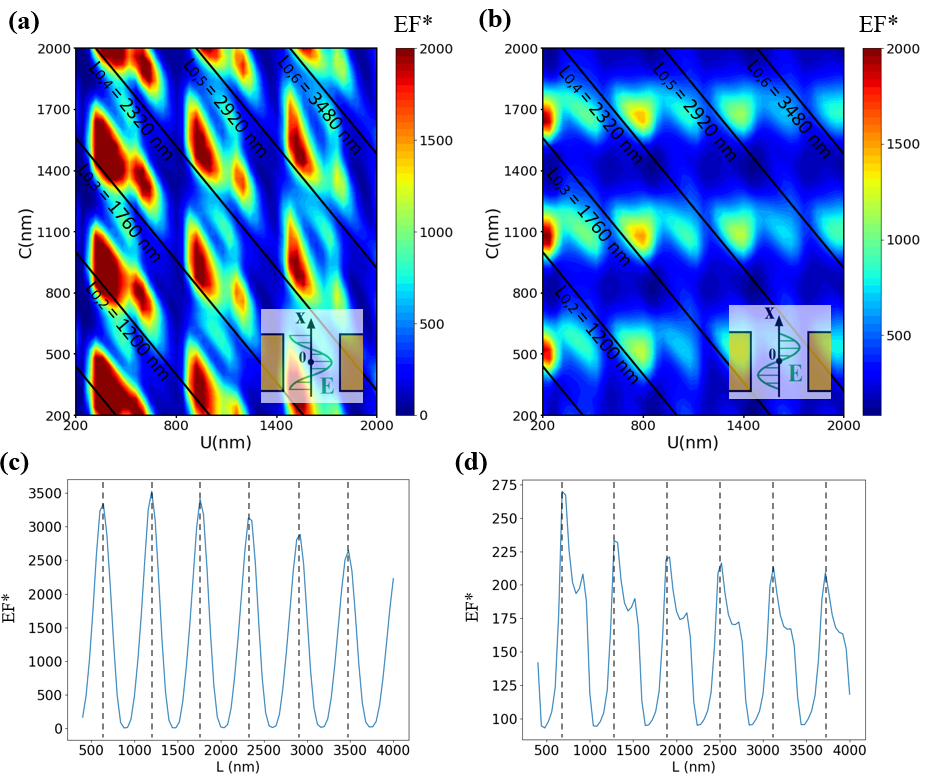}
\caption{\label{fig:5} Modified enhancement-factor colormaps as a function of $U$ and $C$ for $h=200$nm in the presence of the (a) even ((b) odd) combination of plane wave drivings described in the text, i.e., having an antinode (a node) at the center of the gap: $x=0$. Solid lines indicate the $U+C= L_{0,i}$ conditions, i.e., the nanorod's lengths associated to peaks in EF for the uncovered dimer. (c) ((d)) $\mathrm{EF}^{(*)}$ response to the even (odd) excitation combination for the uncovered device as a function of $L=U$ (since $C=0$). Dotted vertical lines show the values of length, $L_{0,i}$, in which the uncovered device reaches an EF peak in presence of only one plane wave direction.}
\end{figure}

Finally, we consider the coexistence of waves with $\uvec{k}=-\uvec{x}$ and $\uvec{k}=+\uvec{x}$.  In order to study limit cases we consider the combinations defined as,
\begin{subequations}
    \begin{align}
        \mathbf{E_{ex}^{even}}&=\mathrm{e}^{-\ci k_0 x}+\mathrm{e}^{\ci k_0 x},\\
        \mathbf{E_{ex}^{odd}}&=\mathrm{e}^{-\ci k_0 x}-\mathrm{e}^{\ci k_0 x}.
    \end{align}
\end{subequations}

Notice that the even case, being proportional to $\cos(k_0 x)$, has an antinode or maximum of the electric field at the center of the gap ($x=0$) whereas the odd case, being proportional to $\sin(k_0 x)$, has a node of the electric field at such position. Giving such positioning of the node or antinode of the electric field one would expect these two conditions to be extreme situations which is convenient for qualitatively demonstrating the largest change to be expected. In presence of a cavity or of a substrate the phase and amplitude of the reflected wave would fall in between producing smaller changes. In order to normalize the electrical field enhancement, we redefined the EF by integrating the average of the fourth power of the incident field, being different in each case, 
\begin{equation}
EF^{(*)}=\frac{\int|\mathbf{E}|^4 dS}{\int|\mathbf{E_{ex}}|^4 dS}.
\end{equation}
In panels (c) and (d) of Fig.\ref{fig:5}  we present the $\mathrm{EF}^{(*)}$ response to such excitation combinations of the uncovered device as a function of $L=U$ since $C=0$. The main peaks of these $\mathrm{EF}^{(*)}$ in both cases fall exactly at the dotted vertical lines signaling the values of $L_{0,i}$, i.e., the lengths developing EF peaks in the uncovered design with only one direction being present. We observe that odd case, having a node at $x=0$ develops a secondary peak at a larger value of $L$. For the covered design, with $h=200$nm as in Fig.\ref{fig:4}, Figure \ref{fig:5}(a) and (b) present the colormaps of $\mathrm{EF}^{(*)}$ as a function of $U$ and $C$ for the even and odd excitation combinations, respectively. We find that for both $\mathbf{E_{ex}^{even}}$ and $\mathbf{E_{ex}^{odd}}$ the main $\mathrm{EF}^{(*)}$ peaks appear for $L$ values larger than the nearest $L_{0,i}$ (lengths developing peaks in the uncovered design). This is an indication that the response to the wave coming from the bottom of the sample that we have introduced in this work (having peaks at lengths larger than $L_{0,i}$ as shown in Figs. \ref{fig:3}(a) and \ref{fig:4}(b)) tends to dominate the behavior in cases of coexistence of both wave directions. A final remark is that, specially in the odd case, there are ranges of $C$ and $U$ in which the obtained $\mathrm{EF}^{(*)}$ is not large. The ranges in which the peaks are expected can be extracted from these plots. However, we conclude that for a particular experimental situation, i.e., substrate or cavity characteristic, it is advised to perform additional calculations in order to avoid a falling of the nanoantenna response into these ranges in which the enhancement factor becomes low.     

\section{Conclusions}
\label{SC:Conclusions}
Given that the covering breaks the top-bottom symmetry of the sample, in this work we have investigated the EF when the excitation comes from the bottom of the sample. The motivation is inspecting the EF behaviour of the nanonantenna in presence of a component of incoming wave from the bottom that can be present due to a reflecting substrate or inside a cavity field. We find that the EF peaks are larger than for the uncovered case in the two excitation cases. The optimal lengths that maximize the EF are shifted, away from the length of the uncovered dimer. In the case of excitation wave coming from the bottom (top) of the sample the optimal length is larger (shorter) than for the uncovered dimer. The results also show that if both waves are simultaneously present, as it would be the case for a highly reflecting substrate, the optimal length tend to be larger than for the uncovered case, i.e., reassembling the situation in which the wave comes from the bottom of the sample. This is valid for both the node or antinode condition lying at the center of the nanorod, with the antinode case being less favourable as the modified EF is found to be reduced. This reduction it is not due to the covered layer but it also affects the uncovered case when it is subject to such excitation. This can be understood as due to the fact that the incident field with a node at the center of the nanorod is not properly matching the shape of the plasmonic mode that generates a large enhancement factor which has a maximum at such position. In summary, we have shown that the covered device develops large EF in several limit cases of the excitation setup. This is very promising because other particular conditions will fall in between the simulated extreme cases. The results also demonstrate that for a given particular setup of substrate (or cavity arrangement) simulations will be required to optimize the generated EF. 

\section*{Conflicts of interest}
There are no conflicts to declare.

\section*{Acknowledgements}
This work has been supported by the Consejo Nacional de Ciencia, Tecnología e Innovación Tecnológica del Perú (CONCYTEC), Contract N° 174-2018-FONDECYT-BM.  A.A.R acknowledges support by PAIDI 2020 Project No. P20-00548 with FEDER funds and additional support from CONICET (Argentina).  M.L.P. acknowledges the funding from FONCyT, PICT-2016-2531, PICT-2020-SERIEA-02705 (Argentina); and Universidad Nacional de Cuyo, SeCTyP, 344 C022 (Argentina). Also, L.M.L.H. acknowledge the hospitality of the Departamento de Física Aplicada II, Universidad de Sevilla, Spain, where part of this work was realized. 

\bibliography{ref.bib}
\end{document}